\documentclass[a4paper,10pt]{aastex}
\usepackage[utf8]{inputenc}
\usepackage{graphicx}
\usepackage{amsmath}
\usepackage{float}

\begin{document}
\shorttitle{Solar toroidal vs. poloidal cycles}
\shortauthors{J. Murak\"ozy}
\title{Phase relationships of solar hemispheric toroidal and poloidal cycles}
\author{J. Murak\"ozy}
\affil{Debrecen Heliophysical Observatory (DHO), Konkoly Observatory, Research Centre for Astronomy and Earth Sciences}
\affil{H-4010 Debrecen P.O.B. 30, H-4010, Hungary}
\email{murakozy.judit@csfk.mta.hu}
\and

\begin{abstract}
The solar northern and southern hemispheres exhibit differences between the intensities and time profiles of the activity cycles. The time variation of these properties has been studied in a previous article on the data of Cycles 12--23. The hemispheric phase lags exhibited a characteristic variation: the leading role has been exchanged between the hemispheres by four cycles. The present work extends the investigation of this variation with the data of Schwabe and Staudacher in Cycles 1--4 and 7--10 as well as Sp\"orer's data in cycle 11. The previously found variation can not be clearly recognized using the data of Staudacher, Schwabe and Sp\"orer. However, it is more interesting that the phase lags of the reversals of the magnetic fields at the poles follow the same variation as that of the hemispheric cycles in Cycles 12-23, {\it i.e.} in four cyles one of the hemispheres leads and the leading role jumps to the opposite hemisphere in the next four cycles. This means that this variation is a long term property of the entire solar dynamo mechanism, both the toroidal and poloidal fields, that hints at an unidentified component of the process responsible for the long term memory.
\end{abstract}


\section{INTRODUCTION}
Several solar phenomena exhibit hemispheric asymmetries and their variations. Most of the relevant papers focus on the variation of the amplitude of the asymmetry. Various periods have been found, 3.7 years \citep{ViBa1990}, periods between 9 and 12 years \citep{Chan2008}, 43.25, 8.65 and 1.44 years \citep{Balletal2005} and a time scale of 12 cycles \citep{Lietal2009}. Signatures of the solar hemispheric asymmetry has been claimed in solar wind speed \citep{ZiMu1998} and the cosmic rays \citep{Krymetal2009}.

\Citet{Balletal2005} have not found an 11 year period in the normalized north-south asymmetry index. This timescale has to be studied in a different way, by examining the phase lags of the hemispheric cycles. Earlier investigations of this behaviour \citep{Wald1957, Wald1971, Zoloetal2009, Lietal2009} indicated that these phase lags exhibit a long term variation.

In our previous paper (\citet{MuLu2012}, henceforth Paper I) the phase lags of hemispheric cycles have been examined in cycles 12--23 by using different methods and a characteristic behaviour has been found: in four consecutive cycles the same hemisphere leads and in the next four consecutive cycles the other hemispheric cycle leads. This characteristic time is reminiscent to that published by \citet{Glei1939}.

The present work was motivated by the question whether this variation was also working before the Greenwich era starting with cycle 12. The other aspect was raised by the recent work of \citet{SvKa2013} examining the hemispheric phase lags of the polarity reversals of the poloidal field. They investigated the time interval of 1945--2011 and this feature can also be examined on a longer time interval in comparison with the phase lags of the hemispheric cycles.

\section{SUNSPOT DATA}
\begin{figure}
 \centering
 \includegraphics[angle=-90, width=14cm]{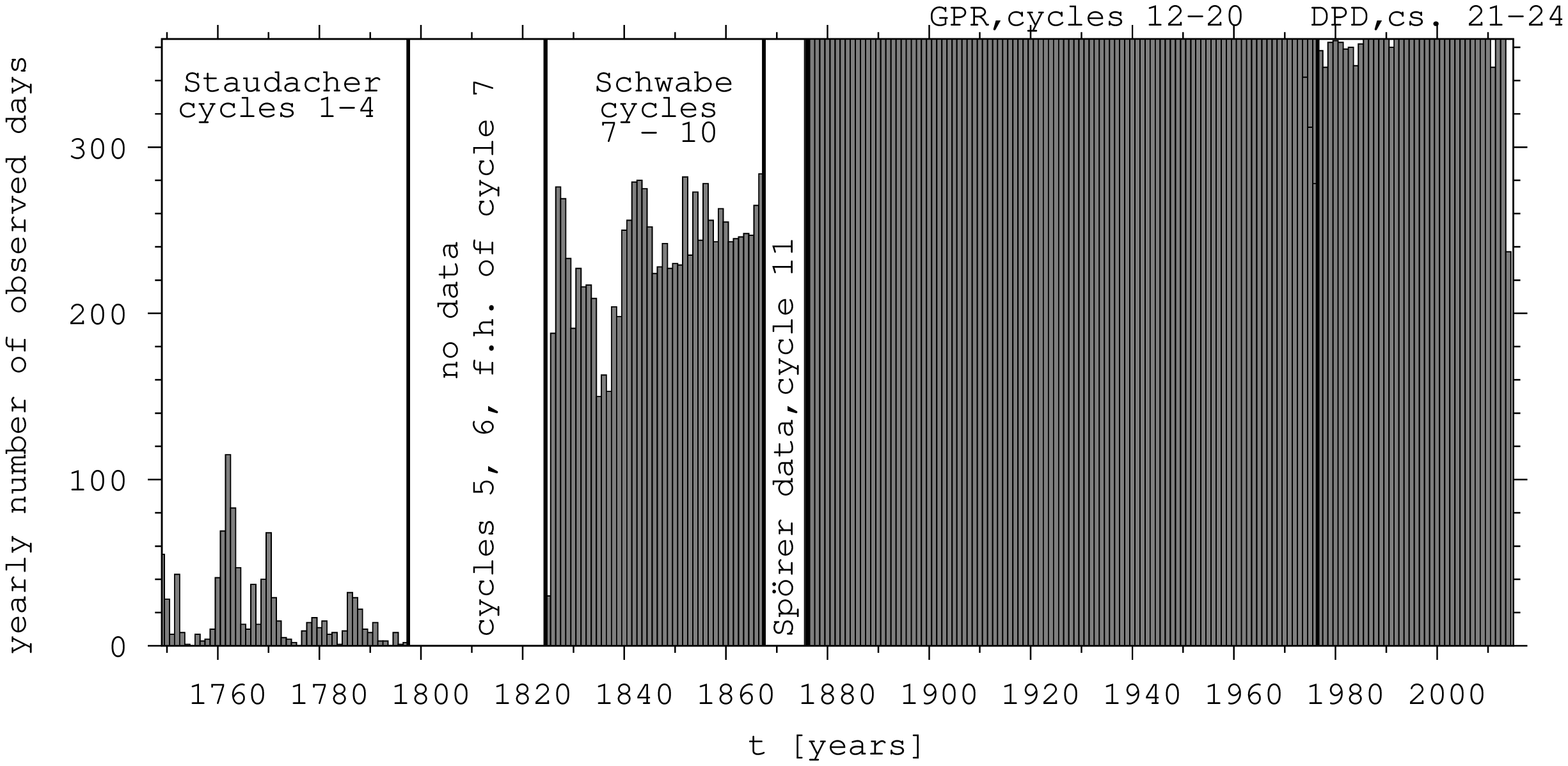}
% \figcaption[elderneweryearly.eps]{Yearly number of observed days. No hemispheric data at all for Cycles 5, 6 and the first half of Cycles 7. Thick vertical lines denote the borders of different databases.\label{spares}}
 \caption{Yearly number of observed days. No hemispheric data at all for Cycles 5, 6 and the first half of Cycles 7. Thick vertical lines denote the borders of different databases.}
 \label{spares}
\end{figure}

The investigation of Paper I was based on the Greenwich Photoheliographic Results, henceforth GPR, \citep{GPR} and the Debrecen Photoheliographic Data, henceforth DPD, \citep{DPD}. For the present extension sunspot data for the previous cycles have been gathered from the observations of Johann Caspar Staudacher for 1749--1798 \citep{Arlt09} and Samuel Heinrich Schwabe for 1825--1867 \citep{Arlt13}. The observations of Staudacher were sparse (Figure~\ref{spares}), in certain years only a few observations were made.

%Figure 1.

After two and a half cycles without any data Schwabe has carried out a long, continuous series of observations covering 43 years. He made observations more  regularly than Staudacher and he identified the sunspot groups, although sometimes he considered two groups as a single one if they were at the same longitude but at different latitudes \citep{Arlt13}. His observations provide position and area data as well as identifying numbers of groups making possible to track the development of certain sunspot groups.

The GPR covers Cycles 12--20, it provides position and area data for sunspot groups, the DPD covers the time interval since Cycle 21 up to now, it contains position and area data for not only sunspot groups but also for all observable individual sunspots. Both catalogues present the data on a daily basis.

As Figure~\ref{spares} shows, hemispheric sunspot data are not available in electronic form for cycle 11, between the Schwabe data and the start of GPR, but this gap can be filled by using the data of Sp\"orer \citep{Spor1874, Spor1878} that have been read into the computer manually. This dataset is based on the observations of Carrington and Sp\"orer between 1854 and 1878 for cycles 10 and 11. Sp\"orer took the observed sunspot groups into account once and weighted them by their area summarizing in five Carrington rotations. He named the obtained data hemispheric frequencies. 
\\Considering the differences between the datasets of different observational periods the input data are somewhat different. In the time intervals of GPR and DPD the monthly sums of sunspot groups are used, in the time intervals of Staudacher's and Schwabe's observations the monthly sums of sunspots and the monthly sums of sunspot areas are considered. No calibrations have been made between these datasets because on the one hand there is no overlap between the datasets of Staudacher and Schwabe, on the other hand all cycles were considered to be separate entities. Only the north-south differences were targeted within each cycle and the strengths of the cycles were not compared to each other. Because of Sp\"orer's weighting method his dataset is not comparable directly with those of Schwabe and the GPR.

The present work uses the monthly values of number of sunspots ($N_\mathrm{{SS}}$) --which is not the well-known International Sunspot Number ({\it ISSN})-- and sunspot group number ($N_\mathrm{G}$)  which consider all sunspots and sunspot groups, respectively as often as they were observable instead of the sunspot group number ({\it SGN}) used in Paper I.

\section{APPLIED METHODS}
\begin{figure}
 \centering
 \includegraphics[angle=-90, width=14cm]{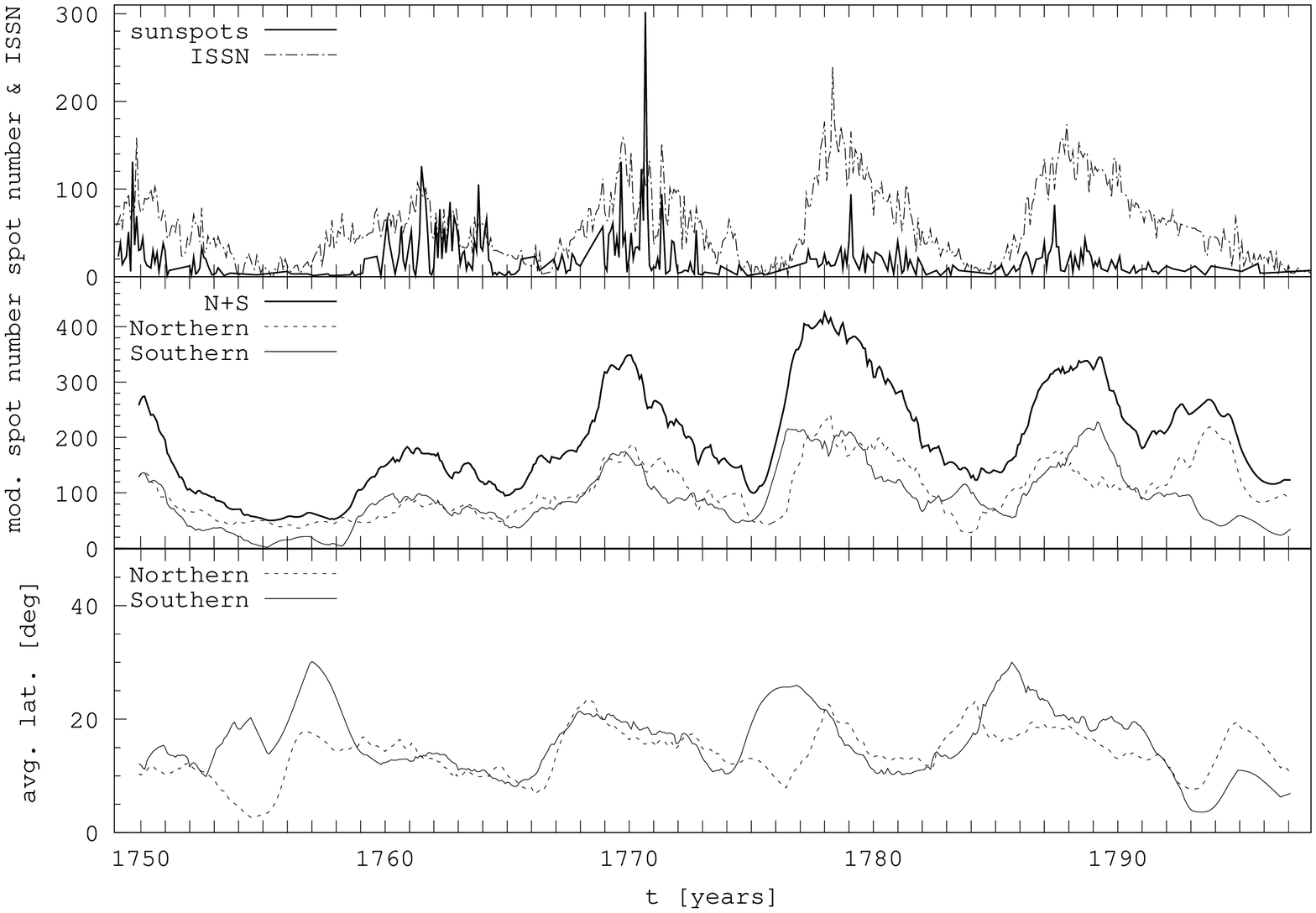}          
 \caption{Profiles of hemispheric cycles in Cycles 1--4 based on Staudacher's observations. First panel: cycles based on the ISSN (dotted line) and the original Staudacher data (solid line). Second panel: cycles based on the modified sunspot number for the whole disk (with thick solid line) and both of the hemispheres. The northern and southern profiles are denoted by dashed and continuous lines, respectively. Third panel: variations of the mean latitudes of hemispheric activity. The profiles are smoothed with 21-month window.}
	\label{recost}
\end{figure}

\begin{figure}
 \centering
 \includegraphics[angle=-90, width=14cm]{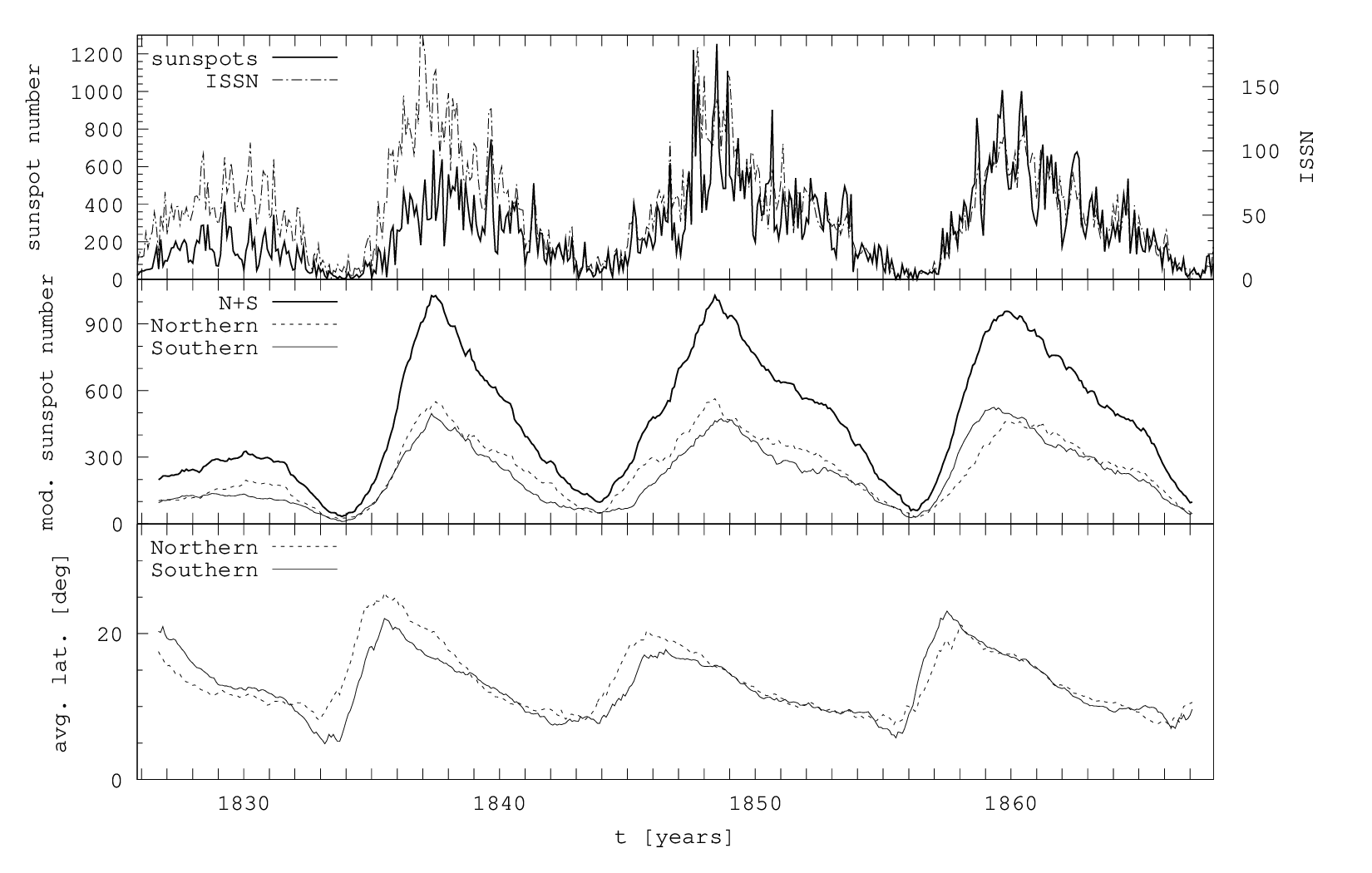}          
 \caption{Profiles of hemispheric cycles in Cycles 7--10 based on Schwabe's observations. First panel: cycles based on the ISSN (dotted line) and the original Schwabe data (solid line). Second panel: cycles based on the modified sunspot number for the whole disk (with thick solid line) and both of the hemispheres. The northern and southern profiles are denoted by dashed and continuous lines respectively. Third panel: variations of the mean latitudes of hemispheric activity. The profiles are smoothed with 21-month window.}
	\label{recosc}
\end{figure}

\begin{figure}
 \centering
 \includegraphics[angle=-90, width=14cm]{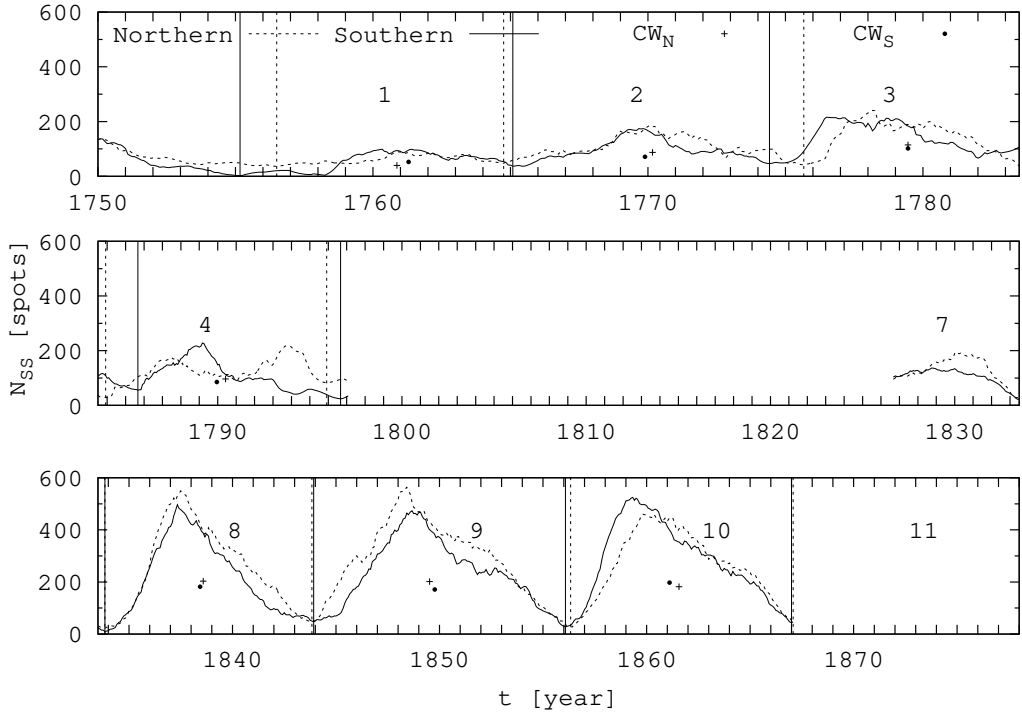}          
 \caption{Profiles of hemispheric cycles in Cycles 1--10, the northern and southern profiles are denoted by dashed and continuous lines respectively. The profiles are plotted by using the monthly number of sunspots ($N_\mathrm{{SS}}$). The profiles are smoothed with 21-month window. Vertical lines denote the times of minima between the global cycles. The centers of weight of the hemispheric cycle profiles have been computed from the unsmoothed profiles, they are denoted by dots and crosses for the northern and southern hemispheres respectively.}
	\label{hemn}
\end{figure}

As is mentioned above, there are missing days in the periods of Staudacher and Schwabe. The approximately true profiles of these cycles can only be reconstructed by using some reasonable substitutions to fill the gaps. The monthly sums of sunspots has been calculated in such a way that the monthly mean value of the observed days was applied for the missing days and these daily values have been summed up for the month (middle panels of Figures~\ref{recost} and \ref{recosc}). As is discernible in the uppermost panels of Figures~\ref{recost} and \ref{recosc} the strengths of these modified cycles fit to the ISSN cycles.
The modified hemispheric $N_\mathrm{{SS}}$ values are plotted in Figure~\ref{hemn}. The hemispheric minima between the cycles are denoted by vertical dashed and solid lines for the northern and southern hemispheres, respectively. The minimum is determined as the time of the lowest monthly value of the inter-cycle profile by smoothing the cycle profiles with a 21-month window. In order to describe each individual cycle as a whole the centers of weight of both hemispheric cycle profiles have been computed by using the original unsmoothed cycle profiles. The positions of the centers of weight ($CW_\mathrm{N}$ and $CW_\mathrm{S}$) are plotted for all hemispheric cycles in Figure~\ref{hemn}, their time difference is the measure of the hemispheric phase lags.

%Figure 2.

%Figure 3.

%\notetoeditor{Figures~\ref{recost} and \ref{recosc} should appear side-by-side in print.}

%Figure 4.

%Figure 5.
\begin{figure}
 \centering
 \includegraphics[angle=-90, width=14cm]{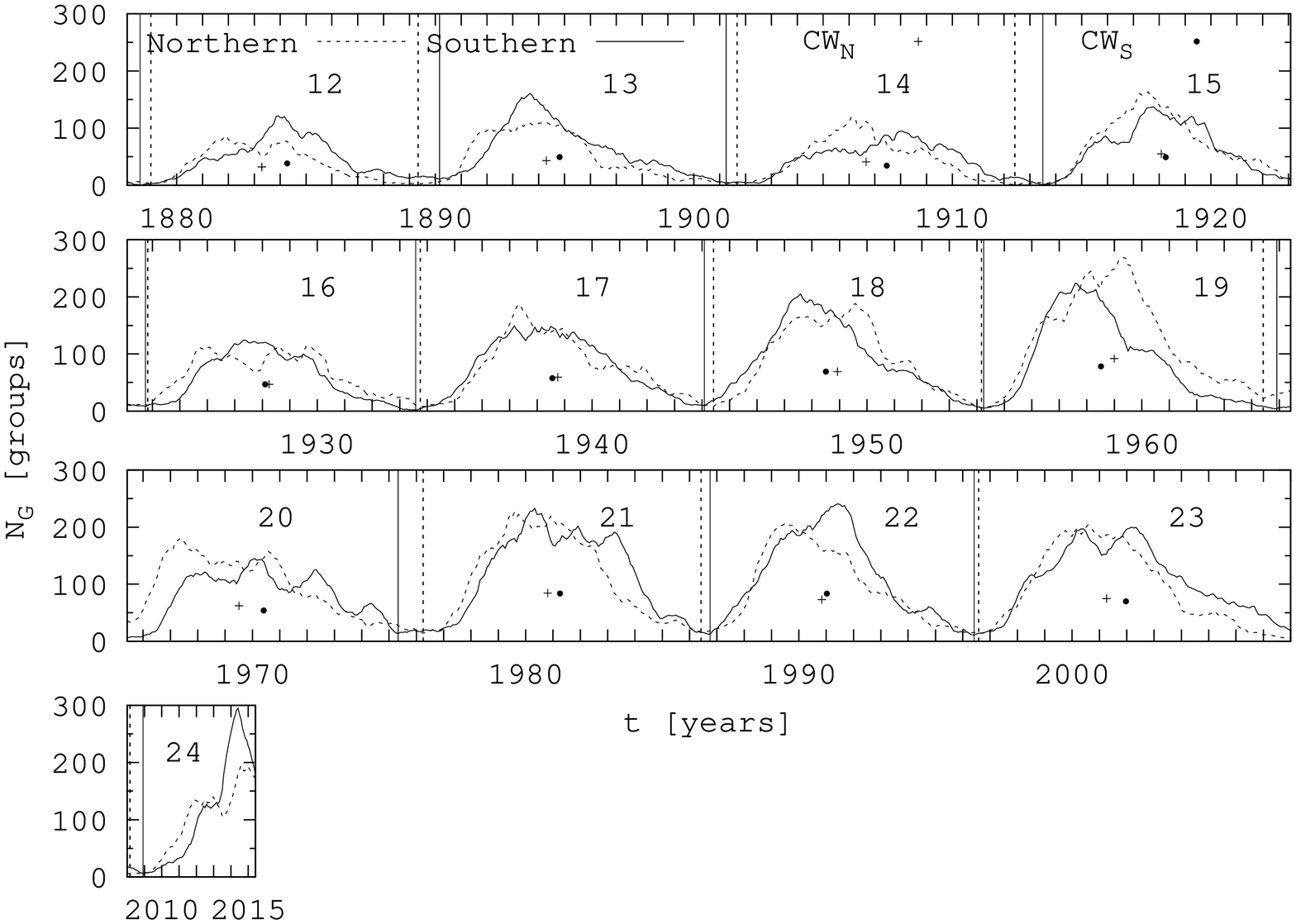}          
 \caption{Profiles of hemispheric cycles in Cycles 12 -- 24 by using the monthly number of sunspot groups. The northern and southern profiles are denoted by dashed and continuous lines respectively. The profiles are smoothed with 11-month window. Vertical lines denote the times of hemispheric minima. The centers of weight of the hemispheric cycle profiles have been computed from the unsmoothed profiles, they are denoted by dots and crosses for the northern and southern hemispheres respectively.}
	\label{hemn24}
\end{figure}

%\notetoeditor{Figures~\ref{hemn} and \ref{hemn24} should appear side-by-side in print.}

Figure~\ref{hemn24} shows the same diagrams for Cycles 12--24, this is the GPR--DPD era. As Cycle 24 is incomplete at the time of this work its center of weight has not been considered. Figure~\ref{hemn24} is similar to Figure 1 of Paper I except the input data. In Paper I the monthly values of sunspot group number ($SGN$) were comupted by counting all the sunspot groups only once in a month. This cannot be done by using the other datasets so that, although an intercalibration cannot be carried out, at least the types of input data can be as consistent as possible.

\begin{figure}
 \centering
 \includegraphics[angle=-90, width=14cm]{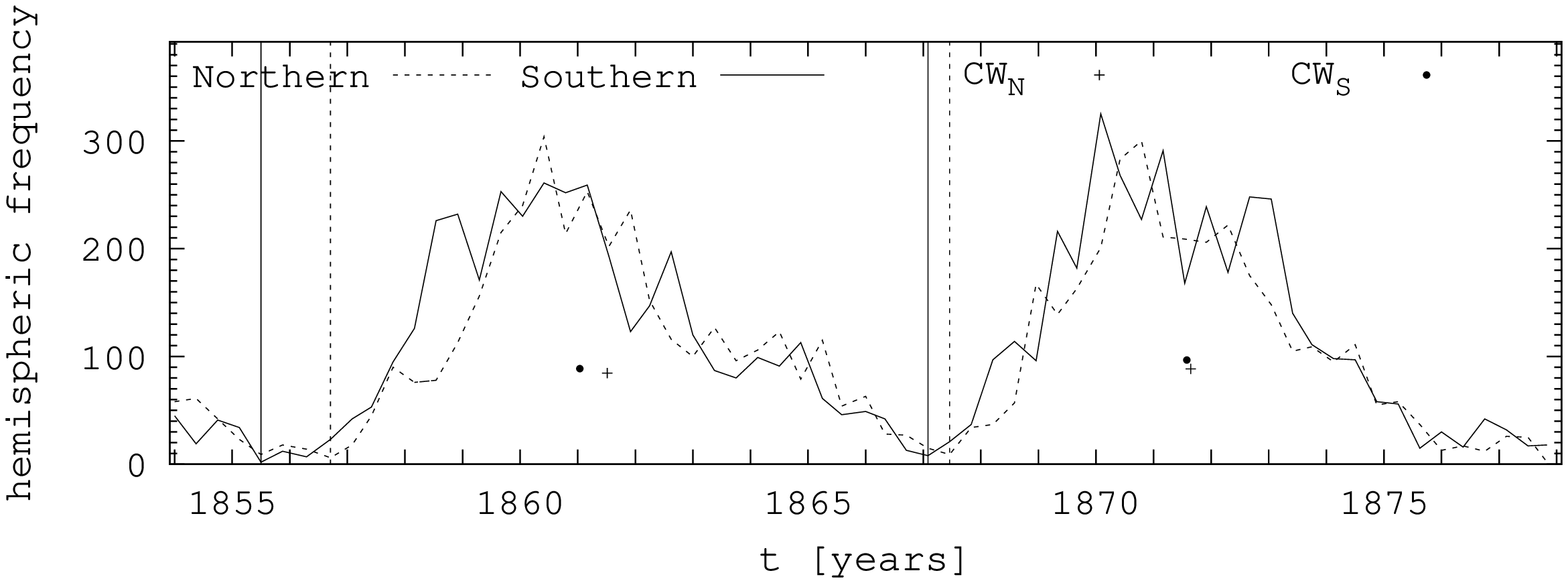}          
 \caption{Profiles of hemispheric Cycles 10 -- 11 by using monthly frequency of sunspot groups provided by Carrington and Sp\"orer and their centers of weight. The vertical lines denote the times of hemispheric minima.}

	\label{hem1011}
\end{figure}

Figure~\ref{hem1011} has been plotted for Cycles 10 and 11 by using Sp\"orer's data. There are no smoothings on these hemispheric profiles because the area weighted hemispheric sunspot data are summarized over five Carrington rotations. The centers of weight are calculated as in the case of $N_\mathrm{{SS}}$ and $N_\mathrm{G}$.

\begin{figure}
 \centering
 \includegraphics[angle=-90, width=14cm]{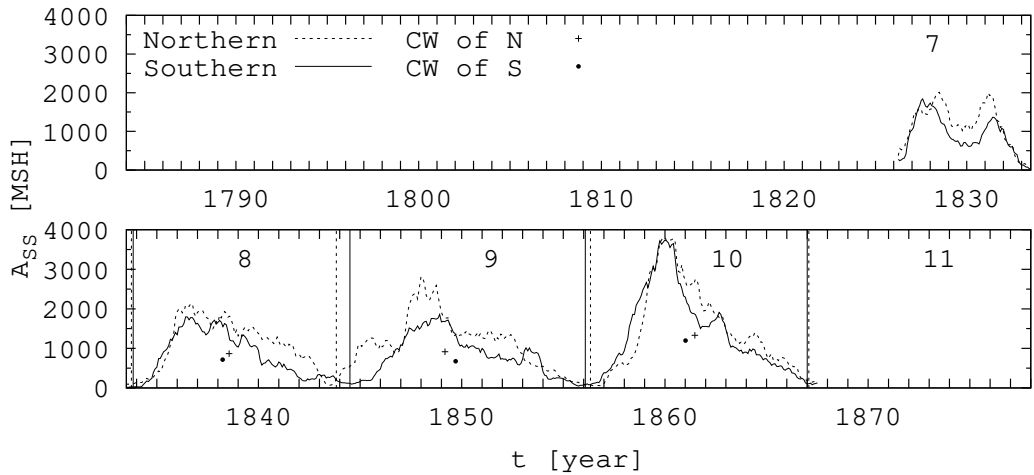}          
 \caption{Profiles of hemispheric Schwabe cycles by using monthly umbral area of sunspots measured in Millionth of Solar Hemisphere (MSH) with 11-month window and their centers of weight (cross/dot means northern/southern value). The vertical lines mean the times of hemispheric minima.}
	\label{hemaS}
\end{figure}

\begin{figure}
 \centering        
 \includegraphics[angle=-90, width=14cm]{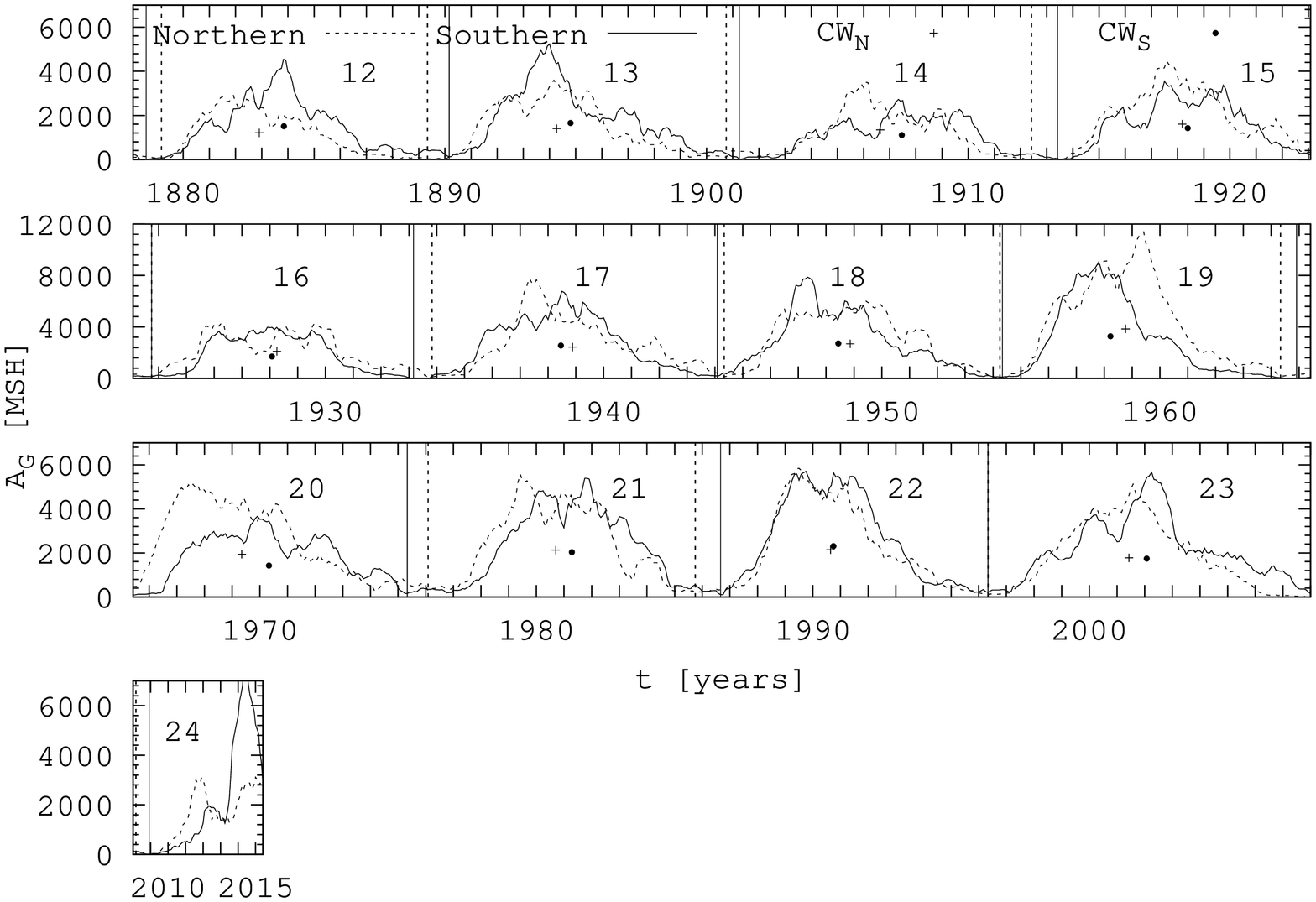}          
 \caption{Profiles of hemispheric cycles  from GPR and DPD by using umbral area of sunspot groups measured in MSH with 11-month smoothing and their centers of weight. The vertical lines mean the times of hemispheric minima.}
	\label{hemaGD}
\end{figure}

To eliminate the uncertainty of determination of sunspot groups the hemispheric monthly umbral area of sunspots ($A_\mathrm{{SS}}$) or sunspot groups ($A_\mathrm{G}$) have been calculated. Figure~\ref{hemaS} shows the $A_\mathrm{{SS}}$ based on the data of Schwabe smoothed with a 21-month window while Figure~\ref{hemaGD} depicts the $A_\mathrm{G}$ by using data of GPR and DPD smoothed with an 11-month window. The hemispheric umbral area is measured in millionth of solar hemispheres (MSH). There is no such a plot for the Staudacher era because his data contain the daily sum of the area of sunspots which does not allow us to distinguish between the hemispheres. 

The centers of weight and the hemispheric phase lags are also determined from the $A_\mathrm{{SS}}$ and $A_\mathrm{G}$ profile, they are plotted in the same way as in the case of $N_\mathrm{G}$ (second panel of Figure~\ref{hists}).

%Figure 6.

%Figure 7.

%Figure 8.

%\notetoeditor{Figures~\ref{hemaS} and \ref{hemaGD} should appear side-by-side in print.}

The long-term variation of the hemispheric phase lags can be studied by different methods. One of them uses the difference between the averages of the normalized asymmetry index in the ascending and descending phases:

\begin{equation}
      \Delta AI=AI_\mathrm{{A}} - AI_\mathrm{{D}} = \\ \left(  {\frac{N_\mathrm{{N}} - N_\mathrm{{S}}}{N_\mathrm{{N}} + N_\mathrm{{S}}}}\right)_\mathrm{A} - \left({\frac{N_\mathrm{{N}} - N_\mathrm{{S}}}{N_\mathrm{{N}} + N_\mathrm{{S}}}}\right)_\mathrm{D}
\label{deltaai}
\end{equation}

\noindent

where $N_\mathrm{N}$ and $N_\mathrm{S}$ are the monthly group numbers of the northern and southern hemispheres respectively, the indices A and D denote the ascending and descending phases respectively. It can be seen that if this difference is positive then the northern hemispheric cycle leads (third panel of Figure~\ref{hists} where the vertical axis is reverted in order to compare them more easily).

Another method is the study of the difference between the maxima of the hemispheric latitudinal distributions of active regions (bottom panel of Figure~\ref{hists}), this method exploits the equatorward shift of the active belt during the solar cycle. This means that the bulge of the latitudinal distribution of the activity is closer to the equator in the leading hemisphere, i.e. $\theta_\mathrm{N}$ -- $\theta_\mathrm{S}$ ($\theta$ denotes the latitude) is negative if the northern hemisphere leads in time. The bars of the bottom panel of Figure~\ref{hists} are calculated by averaging the absolute values of these diferences over the cycles.

\section{HEMISPHERIC PHASE LAGS OVER CENTURIES}
\begin{figure}
 \centering
 \includegraphics[angle=-90,width=8.4cm]{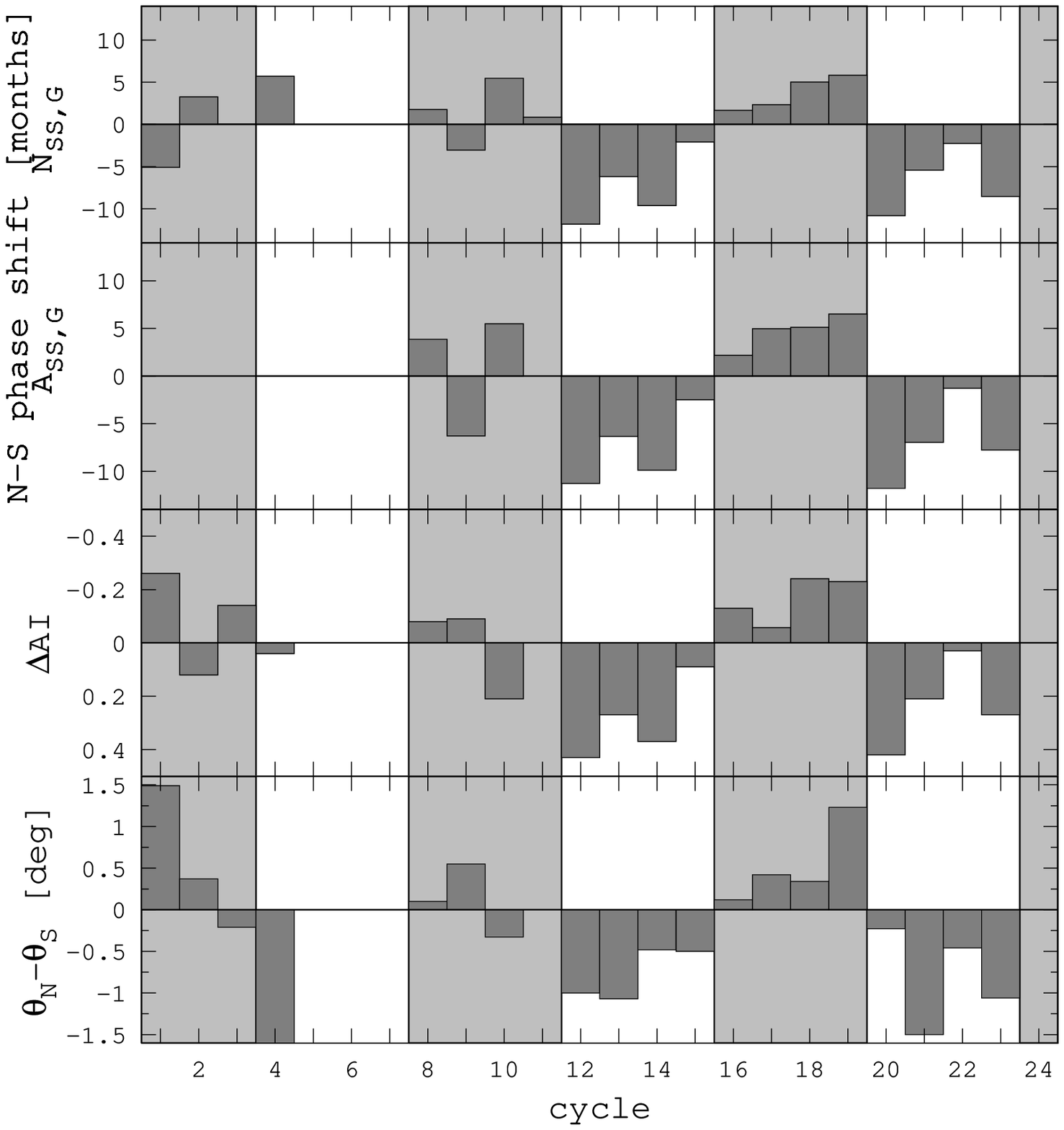}
 \caption{First panel: phase shifts of hemispheric cycles calculated by taking into account the monthly number of sunspot groups and spots. Second panel: phase shifts of hemispheric cycles calculated by the umbral areas of sunspot groups. Third panel: difference between the means of the normalized asymmetry indices on the descending (DP) and ascending (AP) phases (with reversed vertical axis). Fourth panel: hemispheric phase lags obtained from the hemispheric Sp\"orer diagrams. The southern hemisphere leads in those cycles where the columns are positive. In order to better recognize the variation of 4+4 cycles the consecutive groups of four cycles are marked by white and light grey stripes.}
	\label{hists}
\end{figure}

The differences between the time coordinates of the centers of weight of the hemispheric cycle profiles are plotted in the first and second panels of Figure~\ref{hists} by using monthly number of sunspots or sunspot groups and the monthly value of sunspot area, respectively. Similarly to the result in Paper I, the hemispheric phase lags alternating by four cycles are also recognizable during the GPR and DPD era by using the above described four different methods. The case is different during the pre-Greenwich era, beacuse there is no uniform pattern in that period. Cycles 1, 4 and 9 do not fit into the 4+4 alternation by using the first method.

Since the ascending phase of Cycle 7 is missing and the descending phase of Cycle 24 is not complete as yet these cycles are disregarded in these studies. The presented methods do not allow to determine the real hemispheric centers of weight without full coverages of cycles thus these cycles are missing from Figure~\ref{hists}.

Examining the ascending phase of Cycle 24 (Figures~\ref{hemn24} and \ref{hemaGD}) it can be seen that this phase is similar to the case of Cycle 16. The ascending phase of Cycle 24 might indicate northern leading because of the northern predominance of the activity. However, the ascending phase of Cycle 16 also exhibited northern predominance but the examination of the entire hemispheric cycle profiles showed southern leading. This means that the real phase lag can only be determined after the full cycle is completed. \Citet{SvKa2013} also formulated a precautious expectation on this phase lag.

There are no area data for the Staudacher era as described above as well as for the Sp\"orer period. Studying the N-S phase shift by using the area data we can conclude that Cycle 9 here also is an exception to the rule.

The method of asymmetry index gives fairly similar results. When $\Delta AI$ in Equation~\ref{deltaai} is positive/negative then the northern/southern hemisphere leads during the cycle. It can be clearly seen, with reversed vertical axis, in the third panel of Figure~\ref{hists} that this was the case during the Greeenwich and DPD eras. During the pre-Greenwich period Cycles 2 and 10 do not fit into the 4+4 alternation by using this third method.
 
The fourt panel of Figure~\ref{hists} shows the hemispheric phase lags obtained from the hemispheric Sp\"orer diagrams. It can be seen that the 4+4 alternation can be pointed out on the GPR and DPD data, but can not be clearly perceivable in the pre-Greenwich age, because Cycles 3 and 10 are exceptions to the 4+4 alternation.

\begin{table}[htbp]
\begin{center}
  \begin{tabular}{| c | c | c | c | c | c |}
    \hline
    cycle & N-S phase shift $N_\mathrm{{SS,G}}$ & N-S phase shift $A_\mathrm{{SS,G}}$ & $\Delta AI$ & $\Theta_\mathrm{{N}}-\Theta_\mathrm{{S}}$ & authenticity \\ \hline \hline
    1 & -- & no data & + & + & 2/3 \\ \hline
    2 & + & no data & -- & + & 2/3 \\ \hline
    3 & + & no data & + & -- & 2/3 \\ \hline
    4 & -- & no data & + & + & 2/3 \\ \hline
    8 & + & + & + & + & 1 \\ \hline
    9 & -- & -- & + & + & 1/2 \\ \hline
    10 & + & + & -- & -- & 1/2 \\ \hline
    11 & + & no data & no data & no data & 1 \\ \hline
  \end{tabular}
\end{center}
\caption{Authenticities of the cycles. +/-- mean right/wrong results on the basis of Figure~\ref{hists}.}
\label{auth}
\end{table}

It can be discernible that the different methods result in different behavioural patterns in the pre-Greenwich cycles. In order to determine the authenticity of each cycle, let the authenticity of the cycles mean the number of right cases from all the investigatable methods. It can be seen in Table~\ref{auth} created by using Figure~\ref{hists} that there are four cycles with two-thirds, two cycles with a half and another two cycles with one authenticity. In spite of the low coverages of the eight full pre-Greenwich cycles there are six cycles with authenticity higher than a half. Obviously, the results of these kinds of investigations will be better and reliable if the sunspot datasets are full or almost full. This is why the long-term databases are so important.

However, Cycle 10 shows half authenticity in this study by using reconstructed data; it cannot be disregarded that this cycle and Cycle 11 fitted to the 4+4 rule in the work of \cite{Wald1971} on the Z\"urich data (see Paper I).

\section{LONG-TERM VARIATIONS OF POLOIDAL FIELD REVERSALS}
\citet{SvKa2013} investigated the asymmetries of hemispheric activity cycles in connection with the timings of polar field reversals by examining the supersynoptic maps of the Mount Wilson Observatory starting in 1970. They did not study any long-term variations or regularities in this relationship because of the short time interval but the backward extension of the set of the times of polarity reversals makes it possible. Like the long-term sunspot studies and all long-term investigations the work with these data has also to compromise with the broad variety of sources and types of observations.

The most suitable set of dates has been published by \citet{MaSi1986}. Their procedure is based on the method of \citet{McIn1972} who reconstructed the large scale surface magnetic field distribution by using H-alpha synoptic charts. Large regions of radial magnetic fields of opposite polarities are separated by borderlines indicated by filament bands of mainly east-west direction. It is a century old finding \citep{Feny1908} that these filaments migrate toward the poles thus by tracking the poleward migration of these borderlines the time of the polarity reversal can be determined. 

\Citet{MaSi1986} have compiled a set of reversal dates from different sources covering the period 1870--1981. The Kodaikanal H-alpha and Ca II K spectroheliograms cover the period 1904--1964, their reliability in identifying the opposite polarity regions has been checked by comparing them to magnetograms. After 1964 magnetograms were used. The period 1870--1903 has been covered by using the limb filament observations of \citet{Ricc1914}. After 1981 I used the reversal dates published by \citet{SvKa2013} and \citet{SoIn2015} for Cycle 24.

The upper panel of Figure~\ref{polrev} shows the N-S differences between the polarity reversal dates of the poloidal magnetic field while for the sake of comparison the lower panel shows the phase lags of the hemispheric cycles. In those cases (Cycles 12, 14, 16, 19, 20 and 24) when there were two or more polarity changes I have taken into account the dates of the final reversals because the magnetic field can be strongly varying around the time of polarity reversal.

As it can be seen in Table 1 of \citet{MaSi1986} the northern and southern polarity reversals took place simultaneously in Cycles 11 and 13. The time differences are zero in these cycles and these results neither contradict to nor corroborate the examined long-term variation but the pattern of the other cycles is conspicuous. The variation of the poloidal polarity reversals in the upper panel of Figure~\ref{polrev} seems to fit to the regularity of the variation of hemispheric phase lags by 4+4 cycles. Cycle 20 is the only exception to that regularity. 

The similarity of the toroidal and poloidal phase lags is remarkable merely by visual inspection but their comparison is even more informative in Figure~\ref{poltor} showing the diagram of the relationship between these two kinds of phase lags. Two regression lines are indicated, the steeper one disregards the dot of the non-fitting Cycle 20, the less steep line takes it into account. Apparently, the hemispheric poloidal fields sense the statuses of the hemispheric toroidal fields therefore their phase relationsiphs correspond to those of the toroidal fields. This corroborates and generalizes the existence of the phase-lag variation of 4+4 cycles during Cycles 12-23.

\section{DISCUSSION}
The study published in Paper I has been extended in two ways, temporally and physically. 

%Figure 9.

The phase lags of hemispheric cycles have been examined in the pre-Greenwich era, eight additional cycles were more or less suitably covered by the necessary sunspot data. The results show that the phase lag variation by 4+4 cycles can be more or less recognized but with certain exceptions. Therefore it cannot be stated for sure that this variation was working in pre-Greenwich cycles. Either it may have been absent or its existence cannot be pointed out because of the decreasing observational coverage. Otherwise, there are six cycles with two-thirds or more and just two cycles with a half authenticity during the pre-Greenwich times.

An objective physical cause may be the uncertain status of Cycle 4 around 1790, where a cycle may have been lost \citep{Usetal2001}, a statement debated by \citet{Kretal2002}. This cycle, as a single entity, fits unambiguously into the set of 4+4 cycles but the next documented cycle of the group does not. The next group contains two fitting and one unfitting cycle. It cannot be excluded that the case of "missing cycle" temporarily distorted this long-term variation similarly to the Gnevyshev-Ohl rule. As it can be seen in the middle panel of Figure~\ref{recost} the so-called lost cycle can be observable in the northern hemispheric activity by using the modified sunspot data while in the original data of Staudacher (uppermost panel of this figure) can not. The lowermost panel may strengthen the existence of the lost cycle because the mean hemispheric latitudes rise after 1792 as at the beginning of a new cycle and decrease after 1794 however the southern activity continuously decreases after the maximum of Cycle 4.

Anyhow, the Schwabe cycle itself has also stochastic features, even extended solar minima. Any regularities can only work in restricted time intervals.

%Figure 10.
\begin{figure}
 \centering
 \includegraphics[angle=-90,width=8.4cm]{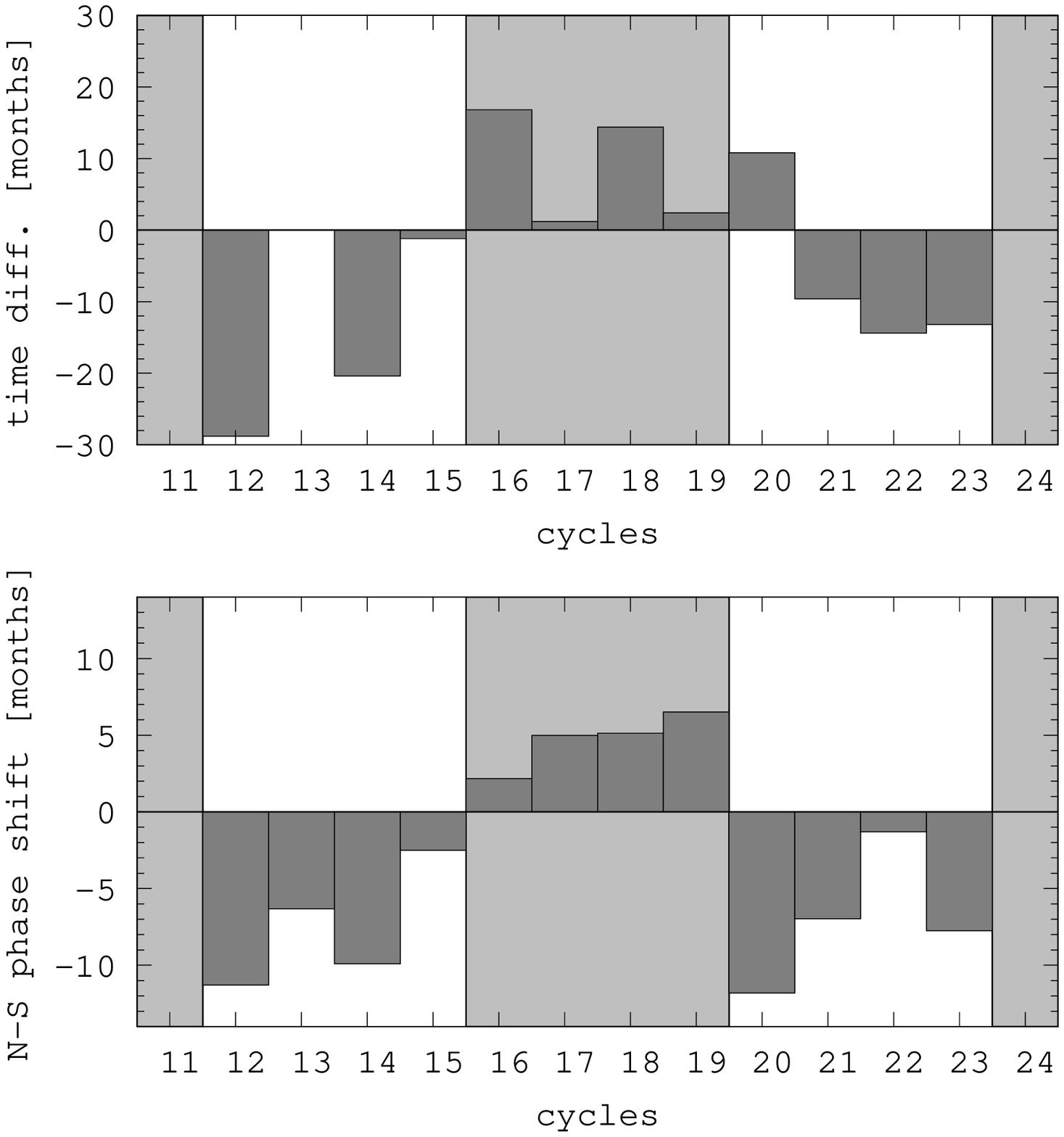}          
 \caption{Upper panel: differences between the dates of hemispheric polarity reversals of the poloidal field. In the cases of Cycles 11 and 13 the polarity reversals happened simultaneously. Positive columns mean that the southern pole leads. Lower panel: phase lags of hemispheric cycles calculated by umbral areas of sunspot groups. White and light grey stripes show a consecutive series of four cycles.}
	\label{polrev}
\end{figure}

%Figure 11.
\begin{figure}
 \centering
 \includegraphics[angle=-90,width=8.4cm]{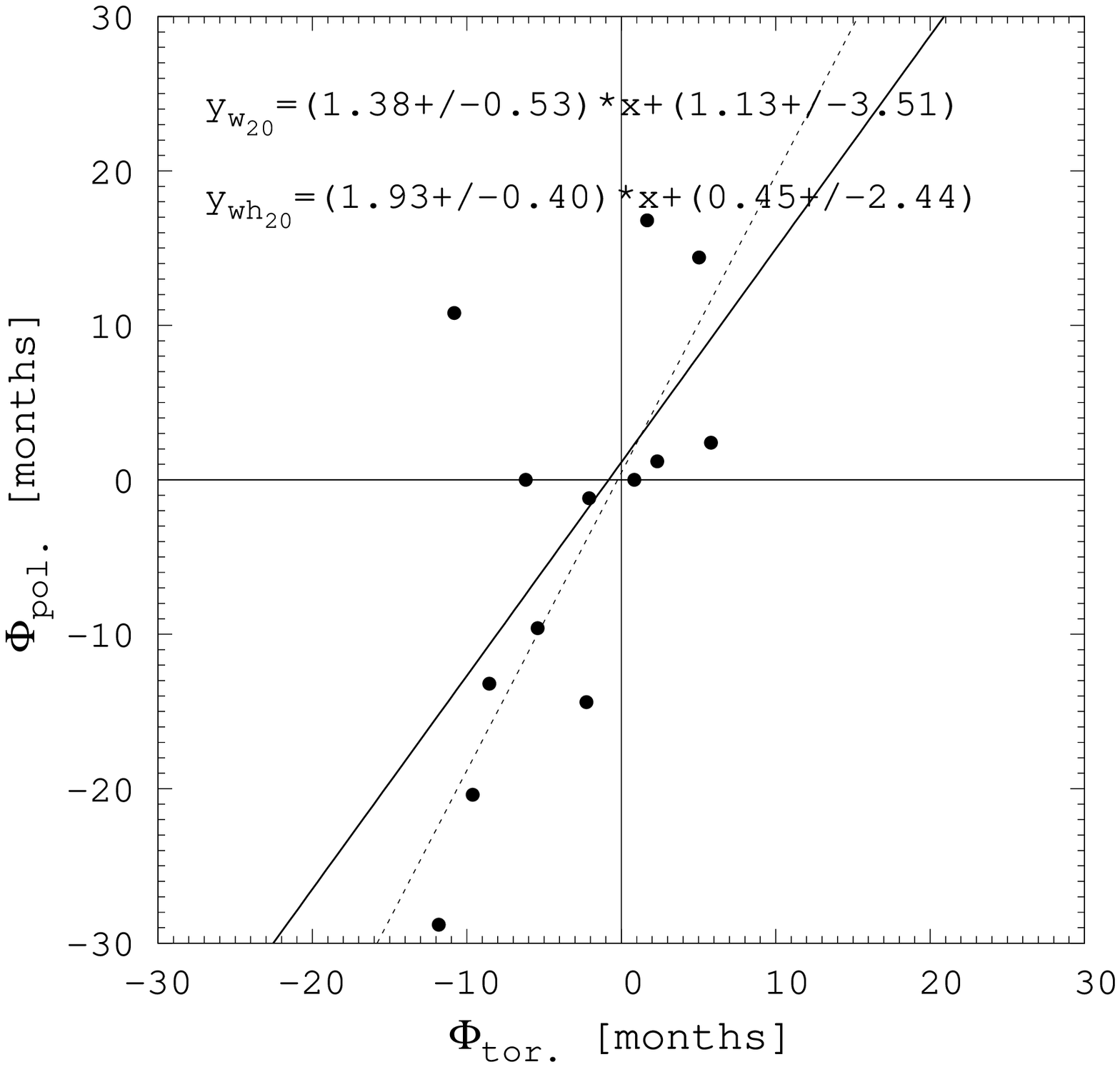}          
 \caption{Differences between the dates of hemispheric polarity reversals of the poloidal field versus phase lags of hemispheric cycles. Two regression lines are fitted. The upper equation(continuous line) takes the non-fitting Cycle 20 into account, its dot is in the upper left quarter of the diagram, the lower equation (dotted line) disregards it.}
	\label{poltor}
\end{figure}

%\notetoeditor{Figures~\ref{polrev} and \ref{poltor} should appear side-by-side in print.}

A more convincing corroboration of the phase lag variations of 4+4 cycles is obtained by the other extension of the study, the examination of the differences between the polarity reversals of the poloidal field on the GPR-DPD era. Figure~\ref{polrev} shows these differences in comparison to the hemispheric phase lags. The two column diagrams are fairly similar with a single exception, Cycle 20. This implies that the regularity of 4+4 cycles in the phase lags is a more general feature of the solar dynamo and involves both the toroidal and poloidal process.

The two topologies are continuously alternating by being transformed into each other but these diagrams may rise a question of "chicken-and-egg"  type. It should be noted  that the columns of the poloidal diagram belong to those of the toroidal one, for instance the phase lag between the reversals of hemispheric poloidal fields denoted by 18 happened around the maximum of Cycle 18. Thus the presented alternation may mean that this specific temporal feature of the solar cycle is ruled by the long term behaviour of the hemispheric toroidal fields. The temporally leading hemispheric cycle is able to initiate an earlier polar reversal than the opposite hemispheric cycle. The presented variation of 4+4 cycles should be the evolutional property of the toroidal field. 

It would be premature to speculate about any underlying mechanisms. Relevant phase  relations have been targeted earlier in different ways. \citet{Stix1976} as well as \citet{ScSt1995} examined theoretically phase relations between $B_\mathrm{\varphi}$ and $B_\mathrm{r}$ fields. 
%\Citet{Zharetal2012} investigated separately the sunspot magnetic fields  (SMF) and the solar background magnetic fields (SBMF), these were considered to represent the toroidal and poloidal fields respectively, they applied principal component analysis and identified waves and their phase relations. \Citet{Shepetal2014} applied the same technique for tracking the SBMF principal component waves and their phase relations and also presented their forecast. These works, however, do not yield clues to the interpretation of the phase lag patterns presented here. 
Apparently, a yet unknown agent has to be identified that might be responsible for this long term behavioural pattern needing long term memory.

\acknowledgments
The research leading to these results has received funding from the European Community's Seventh Framework Programme (FP7/2010-2013) under grant agreement $n^ \circ$ 284461. The Staudacher and Schwabe data are courtesy of Rainer Arlt. Thanks are due to Andr\'as Ludm\'any for reading and discussing the manuscript. The author is deeply indebted to those people for the inspiration who asked the following question ''What guarantees that this variation will be continued before the GPR-era and after Cycle 23?'' in several conversations.

\end{document}